\documentclass[11pt]{article}

\usepackage[preprint]{acl}

\usepackage{times}
\usepackage{latexsym}

\usepackage{microtype}
\usepackage{graphicx}
\usepackage{subfigure}
\usepackage{booktabs} 

\usepackage{hyperref}

\usepackage{nicefrac}       
\usepackage{multirow} 
\usepackage{graphicx}
\usepackage{enumitem}

\usepackage{tabularx}
\usepackage{wrapfig}
\usepackage{algorithm}
\usepackage{algorithmic}

\definecolor{darkblue}{HTML}{2e5f7f}
\definecolor{darkgreen}{HTML}{397d26}

\usepackage{amsmath}
\usepackage{amssymb}
\usepackage{mathtools}
\usepackage{amsthm}
\usepackage{amsfonts}       
\usepackage[capitalize,noabbrev]{cleveref}
\usepackage{pifont}
\usepackage{bbm}

\theoremstyle{plain}

\usepackage{fancyvrb}
\usepackage{fvextra}
\fvset{breaklines=true}
\usepackage{tcolorbox} 
\tcbuselibrary{listings,breakable}
\usepackage{lipsum}
\newtcolorbox{instructionbox}{
  colback=red!5!white,
  colframe=red!75!black,
  fonttitle=\bfseries,
  title=Instruction:,
  breakable
}
\newtcolorbox{answerbox}[1]{
  colback=blue!5!white,
  colframe=blue!75!black,
  fonttitle=\bfseries,
  title=Answer: #1,
  breakable
}
\newtcolorbox{outputbox}[1]{
  colback=green!5!white,
  colframe=green!75!black,
  fonttitle=\bfseries,
  title=Outputs: #1, 
  breakable, 
}
\newtcolorbox{mainbox}[1]{
  colback=black!5!white,
  colframe=black!75!black,
  fonttitle=\bfseries,
  title=#1,
  breakable
}

\usepackage[textwidth=3.5cm,textsize=tiny]{todonotes}

\title{Belief-Aligned Team Formation: \\ Mitigating Cognitive ToM Pitfalls in LLM Multi-Agent Systems}
\author{Jiaqi Shao$^{1,2}$, Tianjun Yuan$^{1}$, Tao Lin$^{3}$, Bing Luo$^{1}$ \\
    $^{1}$Duke Kunshan University, Kunshan, China \\
    $^{2}$Hong Kong University of Science and Technology, Hong Kong, China \\
    $^{3}$School of Engineering, Westlake University, Hangzhou, China \\
    }
\begin{document}
\maketitle

\setlength{\abovedisplayskip}{3pt}
\setlength{\belowdisplayskip}{3pt}
\setlength{\abovedisplayshortskip}{3pt}
\setlength{\belowdisplayshortskip}{3pt}

\begin{abstract}
Theory of Mind (ToM) - the ability to understand others' mental states - is a fundamental aspect of human cognition. 
Recent work has equipped Large Language Model (LLM) agents with ToM capabilities, yet our investigation reveals the pitfalls of ToM in multi-agent systems (MAS): 
agents with ToM abilities often exhibit worse cooperation performance. 
This highlights the challenge of effectively translating ToM abilities into cooperations in LLM-based MAS.
To address this challenge, we propose a ToM-aware team formation mechanism that explicitly considers belief alignment when forming cooperative teams.
Our approach introduces a stable team formation algorithm that optimizes team formation based on both ToM-derived belief alignment and specialized agent capabilities.
Through extensive experiments, we demonstrate 
our mechanism improves the task-specific performance up to 26\% compared to random team formation, while maintaining stable long-term collaboration.
\end{abstract}

\section{Introduction}\label{sec:intro}

Cooperation is a fundamental aspect of multi-agent systems (MAS), enabling agents to work together as a team to achieve common goals and solve complex problems. In recent years, the rapid advancement of large language models (LLMs) has opened up new opportunities for building intelligent MAS. LLMs have demonstrated remarkable capabilities in natural language understanding, generation, and reasoning, such as GPT-family~\cite{eloundou2023gpts}. 
By leveraging these powerful LLMs, MAS can enable agents to communicate and collaborate using natural language, resulting in more flexible and human-like interactions in cooperation tasks.

Recent research in LLM-based MAS has focused on enhancing communication and knowledge sharing among agents~\citep{xu2023exploring, lan2023llm, hua2023war, wu2023chatarena, nascimento2023self, fu_improving_2023}, 
as well as advancing cognitive abilities such as reasoning and Theory of Mind (ToM)~\citep{qi2020adaptive, li2023theory, liang2023encouraging, agashe2023evaluating, lin2024swiftsage}. 
However, despite these advancements, \textit{the interplay between agents' cognitive abilities (like ToM) and cooperation} remains unexamined.

ToM represents a fundamental cognitive capability in multi-agent interactions~\citep{premack1978does}. 
It operates at different levels of complexity, from simple belief attribution to recursive reasoning
``I think you believe I will...". 
Although cognitive depth appears beneficial for team cooperation,
\textit{its impact on cooperation is unclear}. 
Recent studies have shown varying effects of ToM levels on agent interactions~\citep{wang2022tomc,li2023camel,chan2023chateval}, 
suggesting a more nuanced relationship between ToM abilities and cooperation.

\textbf{ToM Capabilities Alone are Insufficient for Effective Cooperation.}
To better understand this relationship, we systematically investigated ToM's impact on cooperation. 
Our investigation reveals: 
\textit{agents with ToM generally exhibit worse performance in cooperation tasks, especially MAS involving high-level ToM agents} 
(detailed in~\cref{sec:motivation}). 
This finding challenges the intuitive assumption that more sophisticated ToM abilities lead to better cooperation. 

In fact, agents with ToM may overthink and anticipate potential conflicts, 
resulting in more cautious cooperation. 
This observation aligns with psychological research~\citep{ridinger2017theory} 
indicating that ToM capabilities alone are insufficient for effective cooperation. 
In addition, \citet{ridinger2017theory} suggests that successful collaboration also requires positive reciprocation and alignment in cooperations.
This finding motivates us to explore ToM-aware coordination mechanisms to enhance agent cooperation.

\textbf{Mitigating ToM Pitfalls.} We propose a novel team formation mechanism that 
explicitly considers belief alignment when forming cooperative teams. 
The belief alignment ensures that agents' beliefs about each other's intentions (derived from ToM) and actions are aligned, 
facilitating reciprocation and cooperation.
Unlike prior ToM applications that focus on agent \textit{capabilities}~\citep{li2023theory}, our approach optimizes \textit{cooperative teams}
that can mitigate the ToM pitfalls of ToM agents while maintaining stable cooperation.

\textbf{Contribution.} We summarized our main contributions as follows:

\begin{enumerate}[nosep, leftmargin=2em]
    \item \textbf{Interplay of ToM and Cooperation:} We explore the relationship between Theory of Mind (ToM) capabilities and cooperative performance in LLM-based multi-agent systems. Our findings indicate that ToM capabilities do not always lead to better cooperation, highlighting the complexity of this interplay.
    
    \item \textbf{Team Formation Mechanism:} We propose a novel mechanism for ToM-aware team formation by forming teams based on ToM-driven belief alignment to encourage reciprocation and cooperation.
    
    \item \textbf{Experimental Evaluation:} We comprehensively evaluate our methods, incorporating ToM capabilities to assess their effectiveness in improving cooperation, team stability, and task-specific performance.
    \end{enumerate}

\section{Related Work}\label{app:detailed_related_work}

\textbf{Agents with Theory of Mind (ToM).} The concept of ToM, which involves the ability to attribute mental states to oneself and others, has been studied extensively in cognitive science and psychology \citep{premack1978does, frith2003development}. In multi-agent reinforcement learning, existing research mainly uses supervised training to equip agents with ToM capabilities for cooperative tasks \citep{wen2018probabilistic, wang2022tomc, oguntola2023theory}.

With the rapid advancement of large language models (LLMs), some recent works have explored empowering LLMs with ToM capabilities \citep{arodi2021textual, li2022pre, zhou_how_2023, xu_opentom_2024}.  
However, these efforts have primarily focused on developing ToM capabilities for individual agents, with limited exploration of how ToM can facilitate cooperation among multiple agents. 
While \citet{li2023theory} studied LLM agents with ToM capabilities and leveraged ToM capabilities specifically for cooperative tasks, the relationship between ToM ability and cooperative performance remains unclear.

In this work, we aim to explore the interplay between ToM ability and cooperation. 
Unlike previous studies that primarily focused on equipping individual agents with ToM capabilities, 
we investigate how different levels of ToM abilities influence cooperation performance in MAS.



\noindent\paragraph{Multi-agent Cooperation.}
Multi-agent cooperation has been studied extensively in both industry and academia, with traditional approaches focusing on game-theoretic frameworks~\citep{yoshida_game_2008} and negotiation and communication strategies \citep{foerster2016learning, Tang_2019_ICCV, yang2024how2comm}, enabling agents to learn cooperative strategies through interaction and information exchange.

The advancements of LLMs have opened up new opportunities for multi-agent cooperation, leveraging their capabilities in natural language understanding, generation, and reasoning \citep{zhang_k-level_2024, talebirad_multi-agent_2023, zhang2024proagent, wu2024your}. 
Recent works have explored utilizing LLM agents to coordinate and cooperate in various tasks and 
frameworks, \textit{e}.\textit{g}., CAMEL \citep{li2023camel} employs role-play to facilitate autonomous cooperation, AutoGen \citep{wu2023autogen} uses conversable agents and conversation programming, and MetaGPT \citep{hong2024metagpt} incorporates human workflows into MAS to ensure more coherent and effective teamwork.

In addition, recent research has focused on improving specific aspects of multi-agent collaboration, such as conversation for knowledge sharing~\citep{xu2023exploring, lan2023llm, hua2023war, wu2023chatarena, nascimento2023self, fu_improving_2023} and cognitive ability~\citep{liang2023encouraging, gong2023mindagent}, demonstrating the potential of leveraging language to facilitate cooperation and coordination. 

For effective cooperation in complex and dynamic environments, some studies focus on the dynamic adjustment of group members and the selection of teammates to improve cooperation and overall performance~\citep{chen2024agentverse, li2023camel, shi2023cooperation}. 
Besides, \citet{liu2024dynamic} proposes the Dynamic LLM-Agent Network (DyLAN) framework to optimize team performance based on task queries and peer ranking. \looseness=-1

Unlike previous research that primarily optimizes team composition, 
our work emphasizes the interaction between ToM abilities and cooperation in MAS, 
particularly in the context of team formation.

\section{Motivation}\label{sec:motivation}

\begin{figure*}[!t]
    \centering
    \includegraphics[width=\textwidth]{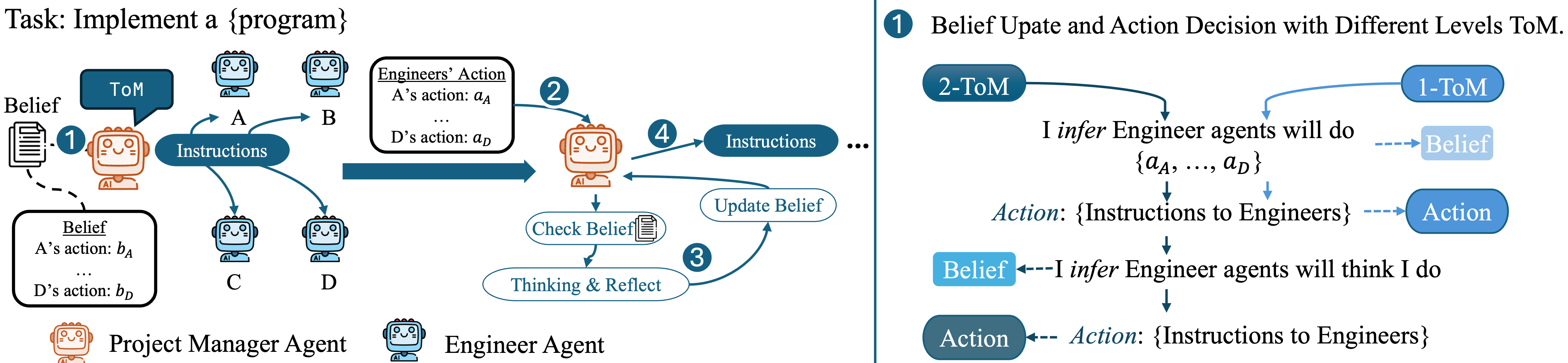}
    \vspace{-1em}
\caption{\small \textbf{Illustration for the MAS with ToM agents}. (\textit{Left}) The MAS for the iterative programming tasks consists of one project manager (\textsf{PM}) with ToM ability and four Engineers. The cooperation involves: \textcolor{darkblue}{\ding{182}}~\textsf{PM} updates its beliefs and takes actions based on its ToM reasoning; \textcolor{darkblue}{\ding{183}}~\textsf{PM} observes Engineers' actions; \textcolor{darkblue}{\ding{184}}~\textsf{PM} think and reflect other's actions and update beliefs; \textcolor{darkblue}{\ding{185}}~\textsf{PM} provides instructions to Engineers. (\textit{Right}) The ToM cognitive thinking process of \textsf{PM} agent involves recursive belief updates and decision-making based on the inferred beliefs. We also employ different ToM settings for the PM and engineers as detailed in~\cref{app:analysis_tom}.} 
\label{fig:motivation}
\end{figure*}

\begin{table*}[bt]
\centering
\caption{\small\textbf{Comparison of performance between agents with lower-level (k=1) and higher-level (k=2) ToM} in the Iterative Programming task on two benchmarks, including \textsc{HumanEval}~\citep{chen2021codex} and \textsc{MBPP}~\citep{austin2021program}, over $\text{R}=1\ldots5$ interaction rounds: High ToM agents show higher initial performance but steeper performance decline over cooperation. Numbers in parentheses show percentage change from R=1 to R=5, with \textbf{bold} indicating high ToM agents with worse cooperation (severer performance degradation).}
\vspace{-0.5em}
\label{tab:iterative_programming_scores_motivation}
\scriptsize
\begin{tabularx}{\textwidth}{lc|Xc|Xc|Xc|Xc|Xc}
\hline
\multicolumn{2}{c}{} & \multicolumn{2}{c}{\texttt{gpt-4o-mini}} & \multicolumn{2}{c}{\texttt{deepseek-V3}} & \multicolumn{2}{c}{\texttt{Llama-3.1-70B-Ins.}} & \multicolumn{2}{c}{\texttt{Doubao-1.5-pro}} & \multicolumn{2}{c}{\texttt{Claude-3.5-sonnet}} \\
&$\text{ToM}^{k}$ & R=1 & R=5 & R=1 & R=5 & R=1 & R=5  & R=1 & R=5  & R=1 & R=5 \\
\hline
\multirow{2}{*}{\textsc{ \textsc{HumanEval} }}  & k=1 & 87.00 & 88.00(+1.15\%) & 94.00 & 94.00(0\%) & 89.13 & {86.17(-3.32\%)} & 94.00 & 92.26(-1.74\%) & 95.88 & {92.00(-4.05\%) }\\
 & k=2 & 91.00 & \textbf{90.00(-1.10\%)} & 96.00 & \textbf{95.00(-1.04\%)} & 92.47 & {90.43(-2.21\%)} & 95.23 & \textbf{93.33(-1.90\%)} & 95.00 & {91.75(-3.42\%)} \\
\hline
\multirow{2}{*}{\textsc{MBPP}} & k=1 & 54.00 & 50.00({-7.41\%}) & 70.00 & {63.00(-10.00\%)} & 49.47 & 34.07({-31.13\%}) & 62.57 & {55.43(-7.14\%)} & 58.33 & 57.29(-1.78\%) \\
& k=2 & 54.00 & \textbf{49.00(-9.26\%)} & 67.00 & 62.00(-7.46\%) & 49.45 & \textbf{32.63(-34.01\%)} & 66.67 & \textbf{57.38(-9.29\%)} & 59.00 & \textbf{52.08(-11.73\%)} \\
\hline
\end{tabularx}
\end{table*}

Recent studies have explored the role of ToM in enhancing agent capabilities and coordination~\cite{li2023theory, yim2024evaluating}. 
While agents equipped with ToM capabilities have demonstrated improved ability to understand and predict others' actions,  
the successful translation of these cognitive abilities into effective cooperative actions remains unexamined. 
To better understand this relationship, we employ a standard ToM model aligned with current research~\citep{zhou_how_2023, street2024llm, xu_opentom_2024}, 
examining how ToM capabilities influence cooperation outcomes in multi-agent systems.

\textbf{ToM Modeling.}
In practice, ToM typically does not exceed second-order reasoning due to cognitive limitations and diminishing returns~\citep{premack1978does, frith2003development}. 
We consider two levels of ToM ability, aligning with existing literature in both human cognitive research~\citep{de2015higher} 
and agent cognitive modeling~\citep{li2023theory}: Low ToM (Level 1), 
where agents can represent and reason about others' beliefs, desires, and intentions; 
and High ToM (Level 2), where agents can additionally consider others' ToM reasoning.

\textbf{Empirical Analysis of ToM in Iterative Programming}
To investigate the relationship between ToM and the performance of MAS,
we conducted simulations of the multi-agent system consisting of one project manager (\textsf{PM}) 
and four engineer agents working together to solve programming tasks, 
as shown in~\cref{fig:motivation} (\textit{Left}). 
Specifically, \textsf{PM} is enabled with ToM ability, 
which means he can recursively infer the actions of other agents (as beliefs) 
and take his own actions accordingly, as illustrated~\cref{fig:motivation}~(\textit{Right}). 
As shown in~\cref{tab:iterative_programming_scores_motivation}, 
\emph{incorporating ToM can introduce side effects that hinder the performance of MAS}. 

Specifically, higher-level ToM agents (k=2) often demonstrate worse cooperation performance than lower-level ToM agents (k=1), 
This suggests that ToM capabilities may lead agents to overthink and anticipate conflicts, resulting in cautious cooperation.
In addition, we also investigate the side effects of incorporating higher ToM in some state-of-the-art methods~\cite{liu2024dynamic,chan2023chateval} in~\cref{app:side_effects}.

\textbf{Organization.}
The rest of this paper is organized as follows:
Section~\ref{sec:formulation} introduces the formal ToM modeling and the team formation problem in multi-agent systems.
Section~\ref{sec:mechanism} describes our proposed team formation algorithm to form cooperative teams among ToM agents. 
Section~\ref{sec:exp} presents the experiments evaluating our proposed mechanism for different cooperation tasks.

\section{Formulation}\label{sec:formulation}
In multi-agent cooperation scenarios, agents typically have defined roles and goals. 
We leverage the alignment between agents' beliefs and actions from ToM to establish teams for team formation. 
In this section, we develop the formal ToM modeling in~\cref{subsec:tom_formulation} and 
the MAS team (team) formation problem in~\cref{subsec:team_matching_formulation}.

\subsection{ToM Formulation for LLM Agents}\label{subsec:tom_formulation}
Theory of Mind refers to the ability to attribute mental states, such as beliefs, intentions, and desires, to oneself and others, and to understand that others may have beliefs 
and intentions that differ from one's own~\cite{premack1978does}.
In the context of MAS, for agent $i$ at round $R$, its $k$-level belief $b_i^{(k)}$ can be formulated as:

\begin{equation}\label{eq:belief_update}
  {\small \mathcal{B}_i^{(k)} = f_{\text{ToM}} \bigg( 
   \underbrace{\mathcal{H}_i^{1:R}}_{\substack{\text{History: observations} \\ \text{\& own actions}}},\,
   \underbrace{\mathcal{A}_{-i}^{1:R-1}}_{\substack{\text{Others' past} \\ \text{actions}}},\,
   \underbrace{\big\{ \widetilde{\mathcal{B}}_{i,m}^{(k-1)} \big\}_{m \neq i}}_{\substack{\text{Predicted beliefs of} \\ \text{others at depth } k-1}}
\bigg) } 
\end{equation}

\vspace{-1em}

Where:
\begin{itemize}[nosep, leftmargin=12pt]
\item $\mathcal{B}_i^{(k)}$ is agent $i$'s $k$-level belief, representing its mental state and prediction about other agents' actions.
\item $\widetilde{\mathcal{B}}_{i,m}^{(k-1)} \!=\! p\big(a_m^R \,|\, \mathcal{H}_m^{1:R}, \mathcal{A}_{-m}^{1:R-1}, \{\widetilde{\mathcal{B}}_{m,l}^{(k-2)}\}_{l \neq m}\big)$ represents agent $i$'s \textit{inference} (not ground truth) about $m$'s beliefs.\looseness=-1
\item Other notations are defined in~\cref{tab:notation}.
\end{itemize}

\begin{table}[t]
  \centering
  \small
  \caption{Key Notation for ToM Formulation}
  \label{tab:notation}
  \vspace{-1em}
  \begin{tabular}{@{}ll@{}}
  \toprule
  Notation & Description \\
  \midrule
  $\mathcal{B}_i^{(k)}$ & Agent $i$'s own $k$-level belief \\
  $\widetilde{\mathcal{B}}_{i,m}^{(k)}$ & $i$'s prediction of $m$'s $k$-level belief \\
  $\mathcal{H}_i^{1:R}$ & $i$'s history: $\{\text{obs } o_i^1, \ldots, o_i^R\} \cup \{\text{own actions}\}$ \\
  $\mathcal{A}_{-i}^{1:R-1}$ & Past actions of all agents except $i$ \\
  $f_{\text{ToM}}$ & LLM-based belief generation function \\
  \bottomrule
  \end{tabular}
\end{table}

\textbf{Base Case ($k=0$).} At zero-order ToM
\(
\mathcal{B}_i^{(0)} = f_{\text{ToM}}\big(\mathcal{H}_i^{1:R}, \mathcal{A}_{-i}^{1:R-1}\big)\,,
\)
capturing naive behavior modeling without recursive reasoning. 

For $k \geq 1$, beliefs nest recursively:
\begin{itemize}[nosep, leftmargin=12pt]
    \item $\mathcal{B}_i^{(1)}$: agent $i$ predicts others’ actions using $k=0$ beliefs.
    \item $\mathcal{B}_i^{(2)}$: agent $i$ predicts others’ $k=1$ beliefs, which themselves depend on others’ predictions of $k=0$ beliefs.
\end{itemize}

\textbf{Chess Example with ToM $k=2$.}
Agent A (White, $\mathcal{B}_A^{(2)}$) observes B's knight move $\mathcal{H}_A^{1:R}$ (board state/history), $\mathcal{A}_{-A}^{1:R-1}$ (B’s past moves) and 
predicts $\widetilde{\mathcal{B}}_{A,B}^{(1)}$ (B’s inferred belief at $k=1$): 
"B expects my queen to retreat."
Then, A uses $f_{\text{ToM}}$ to derive $\mathcal{B}_A^{(2)}$: "B will chase my queen to g4." 

\textbf{\textit{Remarks}}: In LLM agents, $\mathcal{H}_i^{1:R}$, $\mathcal{A}_{-i}^{1:R-1}$, and $\mathcal{B}_i^{(k)}$ manifest as 
natural language interactions~\citep{gur2023real}. 
The ToM hierarchy enables strategic deception that is unattainable with flat reasoning.
We provide carefully designed structured prompts ($f_{\text{ToM}}$) to define each agent's role and level of ToM, guiding the generation of reasoning and facilitating each agent's belief updates~(see \cref{app:prompt_tom} for detailed examples). 

\subsection{Team Formulation}\label{subsec:team_matching_formulation}
This section introduces a team formation mechanism to address: \textit{how to form the stable cooperative team for ToM agents in MAS}.
To begin with, we define the belief-action alignment score used for establishing preference for team formation.
Then, we define the stability condition for team formation, and finally, we introduce the tolerance parameter $\epsilon$ to reduce the search space by filtering out poorly aligned agent pairs.

\paragraph{Team Formation Framework.}
Let $N = \{1,\ldots,n\}$ be the set of agents. A team $\mathcal{T} \subseteq N$ satisfies:  
(a) \textit{Size Constraint}: $|\mathcal{T}| \geq \eta$, where $\eta = \lceil n/2 \rceil$ as default (minimal team size).  
(b) \textit{Preference Consistency}: All agents in $\mathcal{T}$ mutually prefer $\mathcal{T}$ over smaller or alternative teams (no blocking team).

\textbf{Optimal Team Selection.}
The team formation process operates through three components: \textit{team preference}, \textit{stability conditions}, and \textit{welfare optimization}. 

We begin by formalizing how agents evaluate team preference through belief-action alignment metrics.

$\bullet$ \textbf{Team Preference.}  
Each agent $i$ ranks teams $\mathcal{T} \subseteq N$ via its \textit{belief-action alignment score}:
\begin{equation}\label{eq:alignment}
B_i^{(k)}(\mathcal{T}) = \frac{1}{|\mathcal{T}|} \sum_{j \in \mathcal{T}} \phi\Big(\underbrace{\widetilde{\mathcal{B}}_{i,j}^{(k)}}_{\mathclap{\substack{\text{$i$'s prediction of} \\ \text{$j$'s $k$-level belief}}}}, \ \ \ \ \ \ \underbrace{\hat{a}_j}_{\substack{\text{$j$'s actual} \\ \text{action}}}\Big)
\end{equation}
where $\phi(\cdot)$ measures belief-action alignment. 
 Consider two teams $\mathcal{T}_1$ and $\mathcal{T}_2$, if $B_i^{(k)}(\mathcal{T}_1) > B_i^{(k)}(\mathcal{T}_2)$, then agent $i$ prefers (strictly) $\mathcal{T}_1$ over $\mathcal{T}_2$.

For notation simplicity, we use $B_i$ to represent $B_i^{(k)}$ when the context of $k$-level ToM is clear, where $B_i(\mathcal{T})$ captures agent $i$'s evaluation of team $\mathcal{T}$ based on its $k$-level belief predictions about other team members' behaviors.

\textit{\textbf{Remarks}}: 
$\phi(\cdot)$ measures the alignment between beliefs and actions, outputting scores in the range [-1, 1], where 1 indicates perfect alignment, 0 represents neutral/no alignment, and -1 indicates complete misalignment.

To calculate this alignment score, we employ a self-evaluation approach involving prompting the agent to evaluate the alignment between its belief and another agent's action, which is consistent with existing LLM agent literature~\citep{qin2023large,zheng2023judging,liu2024dynamic}.
We also discuss belief-alignment calculation for non-LLM agents in~\cref{app:alignment_example}.

$\bullet$ \textbf{Stability Condition.}
A team $\mathcal{T}$ is stable if:

\begin{equation}
  \nexists\, C \subset \mathcal{T} \colon 
\begin{cases} 
    |C| \geq \eta \\
    B_i(C) > B_i(\mathcal{T})
\end{cases} \quad \forall i \in \mathcal{T}  
\end{equation}

The stability condition ensures no subgroup can defect to form a better-aligned team, 
enforcing mutually beneficial teammates. 
This adapts core concepts from cooperative game theory to ToM agents, 
where preferences are defined based on the alignment between beliefs and actions.

$\bullet$ \textbf{Social Welfare.} 
The \textit{social welfare} of team $\mathcal{T}$ aggregates team preferences $B_i(\mathcal{T})$ defined in~Eq.\ref{eq:alignment}:
\begin{equation}
W(\mathcal{T}) = \sum_{i \in \mathcal{T}} B_i(\mathcal{T}) 
\end{equation}

The optimal team $\mathcal{T}^*$ maximizes social welfare while satisfying stability:
\begin{equation}\label{eq:opt_social_welfare}
\mathcal{T}^* = \text{argmax}_{\mathcal{T} \subseteq N} W(\mathcal{T})
\end{equation}
subject to:
$|\mathcal{T}| \geq \eta$ and no blocking team $C \subset \mathcal{T}$.

To enable tractable team formation, we introduce a tolerance parameter $\epsilon$ that filters agent pairs with insufficient belief-action alignment:
\(
\phi\Big(\widetilde{\mathcal{B}}_{i,j}^{(k)}, \hat{a}_j\Big) \geq \epsilon \,,
\) which \textit{reduces the search space} by eliminating poorly aligned teammates. 
When no pairs satisfy $\epsilon$ (universal misalignment), agents still maintain coherent preferences through $B_i(\mathcal{T})$ in~Eq.\ref{eq:alignment}, enabling fallback team formation based on maximizing social welfare.

\section{ToM-Aware Team Formation}\label{sec:mechanism}
Our mechanism forms belief-aligned teams through welfare optimization with stability constraints.
The cooperative team persists until critical belief-action misalignments trigger reformation, balancing the need for stable collaboration with dynamic mental model updates.

\subsection{Team Formation Mechanism}\label{app:algo} 
The proposed mechanism (\cref{algo:cooperation_matching}) establishes persistent team cooperation through a process of belief-aligned team formation and adaptive reformation. 

Agents initially form team $\mathcal{T} = \mathcal{N} $.
During task execution, agents iteratively refine their beliefs $ \mathcal{B}_i^{(k)} $ through the ToM function $ f_{\text{ToM}} $ (Line~4), which allows them to dynamically update predictions about teammates' behaviors and intentions. This belief updating process reflects the learning and sharing team knowledge.

In addition, the mechanism implements real-time monitoring of belief-action alignment (Line~6).
When alignment falls below threshold $ \phi\Big(\widetilde{\mathcal{B}}_{i,j}^{(k)}, \hat{a}_j\Big)  < \epsilon $, a misalignment counter $c$ tracks potential team instability. 
When the proportion of misaligned pairs exceeds $\theta \cdot |\mathcal{T}|$ (default $\theta = 0.3$) for $\tau$ consecutive rounds (Line~\ref{line:reform}), the reformation initiates re-optimization of $\mathcal{T}$ by maximizing social welfare $W(\mathcal{T})$ (Eq.~\ref{eq:opt_social_welfare}) using updated belief states. 
The $ \tau $-round ($ \tau = 1 $ experimentally) prevents overreaction to temporary fluctuations. 

\begin{algorithm}[t]
  \caption{ToM-aware Team Formation}\label{algo:cooperation_matching}
  \begin{algorithmic}[1]
  \REQUIRE Agent set $\mathcal{N}$; $\epsilon$: alignment tolerance (default: 0.2); $\theta$: stability threshold (default: 30\%)
  \STATE Initialize $\mathcal{T} \gets \mathcal{N}$, $c \gets 0$
  \WHILE{task ongoing}
      \FOR{$i \in \mathcal{T}$}
          \STATE\label{line_belief_update}{Update belief $\mathcal{B}_i^{(k)}$ with Eq.\ref{eq:belief_update}}
          \FOR{$j \in \mathcal{T}$}
              \IF{$\phi\Big(\widetilde{\mathcal{B}}_{i,j}^{(k)}, \hat{a}_j\Big) < \epsilon$ \label{line:alignment_check}} \STATE $c \gets c + 1$ \COMMENT{Track misalignment} \ENDIF
          \ENDFOR
      \ENDFOR
      \IF{$c/|\mathcal{T}| > \theta$  for $\tau$ consecutive rounds}
          \STATE $\mathcal{T} \gets \text{Re-optimize Eq.~\ref{eq:opt_social_welfare} using updated } \{\mathcal{B}_i^{(k)}\}$ \label{line:reform}
          \STATE Reset $c \gets 0$
      \ENDIF
  \ENDWHILE
  \end{algorithmic}
\end{algorithm}

\subsection{Adaptation for Specialized Agent Abilities}

While belief alignment drives stable cooperation, many tasks 
require \textit{domain-specific expertise}. 
We extend the team formation mechanism with skill relevance.  

\textbf{Specialization-Aware Preference Model.}
Let $\alpha_i \in [0,1]$ denote agent $i$'s specialization score 
for specific tasks (e.g., code optimization, medical diagnosis) and
higher values of $\alpha_i$ indicate greater specialized ability for the task. 

We reformulate the team preference $B_i(\mathcal{T})$, defined in~Eq.\ref{eq:alignment}, as $B'_i(\mathcal{T})$:
\[
B'_i(\mathcal{T}) = B_i(\mathcal{T}) + \lambda \cdot \frac{1}{|\mathcal{T}|} \sum_{j \in \mathcal{T}} \alpha_j\,,
\]
where $\lambda$ balances alignment vs. expertise (default $\lambda=1$). 


By incorporating the specialized ability scores $\alpha_i$ into the team preference, 
our team formation algorithm will prioritize agents 
with higher specialized skills for specific tasks and consider the belief alignment. 
This adaptation ensures that agents with crucial specialized abilities 
are more likely to be included in the formed cooperative team.

{\textit{\textbf{Remarks}}: This formulation maintains transitivity in 
team preferences while balancing belief alignment ($B_i(\mathcal{T})$) with task-specific capabilities ($\{\alpha_i\}_{i \in \mathcal{T}}$). 
The teams now form based on both belief-action alignment and specialized abilities. In practice, agents can compute $\alpha_i$ via self-report or skill tests.
We conduct case study incorporating specialized ability scores (detailed in~\cref{app:specialized_ability}) to enhance team performance for tasks requiring specific technical expertise.


\begin{table*}[th]
  \centering
  \caption{\textbf{Comparison of cooperation performance} (Pass@1) between agents with lower-level (1-level) and high-order (2-level) ToM in the Iterative Programming task (\textsc{MBPP}) over five interaction rounds: Random team formation (baseline), both ToM levels show comparable performance; Our proposed team formation improves the cooperative task performance.}
\label{tab:iterative_programming_scores_mbpp}
  \vskip 0.15in
  \scriptsize
  \begin{tabular}{c|cc|cc|cc|cc|cc}
  \hline
  \multirow{2}{*}{Setting} & \multicolumn{2}{c|}{\texttt{deepseek-V3}} & \multicolumn{2}{c|}{\texttt{llama-3.1-70b-Ins.}} & \multicolumn{2}{c|}{\texttt{claude-3.5-sonnet}} & \multicolumn{2}{c|}{\texttt{gpt-4o-mini}} & \multicolumn{2}{c}{\texttt{Doubao-1.5-pro}} \\
  & 1-ToM & 2-ToM & 1-ToM & 2-ToM & 1-ToM & 2-ToM & 1-ToM & 2-ToM & 1-ToM & 2-ToM \\
  \hline
  Beaseline & 63.00 & 62.00 & 34.07 & 32.63 & 57.29 & 52.08 & 50.00 & 49.00 & 55.43 & {57.38} \\
  Ours & \textbf{64.00} & \textbf{65.00} & \textbf{37.76} & \textbf{38.54} & \textbf{57.58} & \textbf{57.00} & \textbf{60.00} & \textbf{61.62} & \textbf{57.58} & \textbf{64.80} \\
  \hline
  \end{tabular}
\end{table*}
\begin{figure*}[t]
\centering
\includegraphics[width=.9\textwidth]{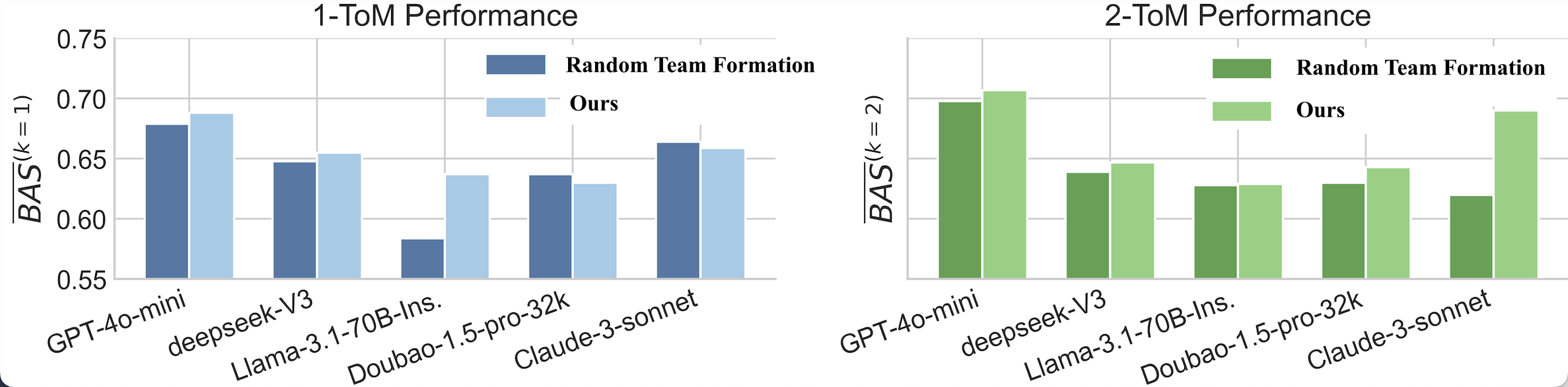}
\caption{\textbf{Comparative analysis of belief alignment across ToM levels} in (\textsc{ \textsc{MBPP} }) at round 5 under ``with random team formation" and ``our proposed team formation" settings.}
\vspace{-10pt}
\label{fig:bas_comparison}
\end{figure*}

\section{Experiment}\label{sec:exp}

\textbf{Experiment Setup.}

The experiment setup involves extending the MetaGPT framework~\citep{hong2024metagpt} to incorporate the proposed multi-agent LLM cooperation mechanism. We vary different ToM configurations for MAS, including 1-level, and 2-level ToM for different LLM agents.
We evaluate our proposed mechanism on the following cooperative tasks: \looseness=-1
\begin{itemize}[nosep, leftmargin=12pt]
\item \emph{\textit{Iterative Programming:}}
In this task, agents simulate a software development team, with each agent assuming different roles. We use \textsc{ \textsc{HumanEval} }~\citep{chen2021codex} and \textsc{MBPP}~\citep{austin2021program} as benchmarks to evaluate the performance of ToM agents in this multi-agent environment.
\item \emph{\textit{Debate:}}
In a debate setting, agents are divided into two sides (affirmative and negative) and engage in a multi-round debate on a given topic. The agents collaborate within their respective sides to present arguments and counterarguments.
\item \emph{\textit{Logical and General Reasoning:}}
We evaluate the effectiveness of our team formation mechanism in reasoning tasks using the AQUA-RAT dataset~\citep[Logic Reasoning]{ling2017program} and MMLU dataset~\citep[General Reasoning]{hendryckstest2021}, where agents collaborate to solve complex reasoning questions. The baselines are ChatEval and DyLAN with ToM agents.
\end{itemize}

\textbf{Models.} 
We utilize five state-of-the-art LLMs in our multi-agent cooperation scenarios: 
\texttt{gpt-4o-mini}~\citep{OpenAI2023GPT}, \texttt{deepseek-V3}~\citep{deepseek2024}, \texttt{llama-3.1-70b-Instruct}~\citep{Meta2024Llama}, \texttt{doubao-1.5-pro-32k}~\citep{doubao2024}, and \texttt{claude-3.5-sonnet}~\citep{Templeton2024Scaling}.

\textbf{Evaluation Metrics.}
In addition to task-specific performance metrics, such as the pass rate for programming, we define two core metrics to evaluate team stability and belief alignment under ToM reasoning. 
These metrics are aligned with our formulation (\cref{sec:formulation}) and our algorithm.

$\bullet$ \textit{Belief Alignment Score.}  
For agent \(i\) with ToM level \(k\), $B_i^{(k)}$ measures the mean alignment between predicted beliefs \(\widetilde{\mathcal{B}}_{i,j}^{(k)}\) and observed actions \(\hat{a}_j\) for teammate $j$ as defined in~Eq.\ref{eq:alignment}.
We consider the team-level belief alignment for team $S$ can be computed as $\overline{\text{BAS}}^{(k)} = 1/|S| \cdot \text{B}_i^{(k)}$.

$\bullet$ \textit{Team Stability Score.}  
 We measure the stability of teams formed by agents with ToM capabilities by considering the average lifetime of these teams. 
The team lifetime is defined as the number of cooperation rounds that a team remains stable before a reformnation occurs due to belief-action misalignment exceeding a predefined tolerance threshold $\epsilon$. 
To calculate the average team lifetime, we use $\frac{1}{m} \sum_{c=1}^m l_c$, where $m$ represents the number of times reformation is triggered throughout the cooperation rounds, and $l_c$ denotes the lifetime of the teams formed after the $c$-th team formnation.

\begin{table*}[t]
  \small
  \caption{\textbf{Comparison of the team stability and Pass@1 performance of MetaGPT and our proposed approach} with 1-ToM and 2-ToM agents using our team formation mechanism in the Iterative Programming task. The team stability is measured by the average number of rounds the formed teams remain stable out of the total 5 rounds.}
  \label{tab:metagpt_scores}
  \centering
  \vskip 0.15in
  \begin{tabular}{r|c|c||c|c}
  \hline
  \multicolumn{1}{c}{} & \multicolumn{2}{c}{\textsc{ \textsc{HumanEval} } } & \multicolumn{2}{c}{\textsc{MBPP} } \\
  & Team Stability & Pass@1  & Team Stability & Pass@1\\
  \hline
  MetaGPT & -- & 85.4\% &--& 86.5\% \\
  1-ToM (Ours) & 3.4/5 & 87.2\% & 3.7/5 & 88.2\%\\
  2-ToM (Ours) & \textbf{3.6/5} & \textbf{90.0}\% & \textbf{4.0/5} & \textbf{90.4}\% \\
  \hline
  \end{tabular}
  \vspace{-1em}
  \end{table*}

\subsection{Main Results}\label{exp:main_results}

To evaluate the influence of ToM on cooperation in MAS, 
we conducted experiments on the system consisting of one Project Manager (\textsf{PM}) and four \textsf{Engineers} tasked with iterative programming. 
We varied the \textit{\textsf{PM}'s ToM ability} between \textit{1-level (low)} and \textit{2-level (high)} 
to observe its effects on team formation and task performance (more evaluations with different ToM configurations are provided in~\cref{app:analysis_tom}).
The proposed team formation mechanism was used to form the cooperative team based on belief-action alignment, 
while the baseline setting uses random team formation (with the same team size $\lceil n/2 \rceil$) was also evaluated for comparison.

\noindent\textbf{Task Performance:}
\cref{tab:iterative_programming_scores} reveals that introducing the our proposed mechanism~(detailed in~\cref{app:algo}, \cref{algo:cooperation_matching}) improves cooperation for both low and high ToM agents. 
Notably, \textit{high ToM agents demonstrate better task performance compared to low ToM agents}. 
This highlights that our team formation mechanism effectively enables high ToM agents's potential to utilize their ToM abilities for long-term cooperation. 
Similar results for the \textsc{HumanEval} are provided in \cref{app:exp}.

\noindent\textbf{Belief Alignment: }
\cref{fig:bas_comparison} shows the team belief alignment scores $\overline{\text{BAS}}^{(k)}$ of 1-ToM and 2-ToM agents with random team formation and our proposed mechanism. 
The results demonstrate that our team formation mechanism generally \textit{improves belief alignment} - teams achieve higher $\overline{\text{BAS}}^{(k)}$ than those with random team formation, with improvements of up to 11\% for 2-ToM agents. 
The only exceptions were 1-ToM agents using doubao-1.5-pro-32k and claude-3.5-sonnet, which showed comparable alignment scores with and without our proposed team formation method.
This improvement is particularly pronounced for 2-ToM agents, suggesting that the welfare-optimized team formation enables more sophisticated ToM reasoning to translate into effective cooperation.
The consistently higher alignment scores also indicate that our mechanism successfully \textit{encourages reciprocal cooperation by forming teams whose beliefs and actions are well-aligned}.

\textbf{Team Stability:}
To assess team stability and task performance, we compared our approach with MetaGPT using 1-ToM and 2-ToM agents on both benchmarks. 
Task performance was measured using the Pass@1 metric~\citep{hong2024metagpt},
 which evaluates the percentage of test cases passed by generated code on its first attempt. 
 Team stability was quantified as the average number of rounds (out of 5) that teams remained stable, 
 where a team is stable if no agent prefers forming a new team.
As shown in \cref{tab:metagpt_scores}, 
our approach outperforms MetaGPT in both task performance and team stability. 
For example, \textit{2-ToM agents with proposed team formation achieve the highest Pass@1 scores of 90.0\% (\textsc{HumanEval}) and 90.4\% (\textsc{MBPP})}. 
They also maintain stable teams for an average of \textit{3.6 out of 5 rounds on \textsc{HumanEval}} and \textit{4.0 out of 5 rounds on \textsc{MBPP}}. 
These results underscore that our proposed mechanism enables ToM agents to form more stable teams and deliver better task outcomes. 
By aligning agents' beliefs and actions, 
our method \textit{fosters long-term team stability and enhances overall collaboration effectiveness}.

\subsection{Case Study: Debate}\label{exp:case}

To evaluate the effectiveness of our proposed team formation mechanism, we designed a debate task using the MetaGPT framework~\citep{hong2024metagpt}. In this task, agents (\texttt{gpt-4-0613}) are divided into two sides: the affirmative side and the negative side, with each side comprising three debater agents. Each side's objective is to cooperate and win the debate.\\ 
At the beginning of our investigation, we explore the cooperation performance of low and high ToM agents in a debate setting with the topic "Should the Death Penalty be abolished?"~\citep{roush2020debatesum, baturo2017understanding}, and both low and high ToM agents are assigned to the affirmative side.
The following example presents the thinking and actions of debaters with different levels of ToM, as detailed in~\cref{app:debate}. 
We observe that higher ToM agent acts as less cooperative, which confirms our previous analysis.
\begin{mainbox}{Compare Cooperation Performance between low and high ToM agents}
\small
\textbf{Thinking and Action of 1-ToM Debater}: 
I anticipate my teammates making these common arguments against the death penalty. \textit{To complement their points}, I will focus my argument on the lack of evidence supporting the death penalty's effectiveness as a crime deterrent. 
I will argue that there is no conclusive evidence showing that the death penalty deters crime more effectively than lengthy imprisonment, making it an unnecessary and unjustified punishment.  \textcolor{blue}{$\Leftarrow$ \texttt{Focuses on complementing their teammates' points \& Good Cooperation}}

\textbf{Thinking and Action of 2-ToM Debater}:
My arguments about deterrence and discrimination will prompt my teammates to make related points to strengthen the overall case. \textit{I should use my next arguments to introduce new issues}.
I will argue that the death penalty often involves inhumane and botched executions, and that most democracies and U.S. states have abolished or prohibited it. \textcolor{red}{$\Leftarrow$ \texttt{Overthinking \& Worse Cooperation}}
\end{mainbox}

\begin{table*}[t!]
  \centering
  \small
  \caption{\textbf{Comparative Evaluations for Logic Problem-Solving and General Reasoning}: The baselines ChatEval and DyLAN involves ToM agents with random team formation.}
  \label{tab:comparative_evals}
  \vspace{-1em}
  \begin{tabular}{lcccc}
  \toprule
   & \multicolumn{2}{c}{\textbf{Logic Problem-Solving}} & \multicolumn{2}{c}{\textbf{General Reasoning}} \\
   \cmidrule(lr){2-3} \cmidrule(lr){4-5}
   & \texttt{gpt-3.5-turbo} & \texttt{gpt-4o-mini}& \texttt{gpt-3.5-turbo} & \texttt{gpt-4o-mini}\\
   & Acc (\%) & Acc (\%) & Acc (\%) & Acc (\%) \\
  \midrule
  ChatEval w. ToM & 40.23 & 69.14 & 54.39 & 69.90 \\
  DyLAN w. ToM & 43.50 & 68.50 & 57.92 & 72.98 \\
  Ours & \textbf{45.70} & \textbf{75.39} & \textbf{60.94} & \textbf{75.57} \\
  \bottomrule
  \end{tabular}
  \vspace{-1.5em}
  \end{table*}
  
Then, 
we conducted experiments under three different settings to show the effectiveness of our team formation mechanism for enhancing cooperation: 
(1) \textit{No-ToM (Baseline)}: Each side consists of 3 debater agents without ToM. For each speech, two debaters were randomly selected from each side; 
(2) \textit{ToM with random team formation}: The three debaters on the affirmative side have varying ToM levels~(0-level, 1-level, and 2-level), while the negative side had no ToM. For each speech, two debaters were \textit{randomly selected} from each side. This setting allows us to evaluate the impact of having an affirmative team with ToM against a team without ToM; 
(3) \textit{ToM with proposed team formation}: Similar to the ToM with random team formation, the affirmative side has three debaters with varying ToM levels. However, the affirmative side selects two debaters using our proposed team formation mechanism.

\begin{table}[t]
\centering
\caption{\textbf{Win rates for both debate sides}. The ToM (baseline) utilizes random team formation.}
\label{tab:experiment_results}
\vspace{-1em}
\begin{tabular}{r|cc}
\hline
\textbf{Setting} & \textbf{Aff.} & {\textbf{Neg.}} \\
\hline
No-ToM & 65.45\% & 34.55\% \\
ToM (baseline) & 61.82\%  & 25.45\% \\
Ours & \textbf{67.27\%} & \textbf{36.36\%} \\
\hline
\end{tabular}
\end{table}

\textbf{Enhance Team Formation for ToM agents}. For the debate topic ``Should the Death Penalty be abolished?", the debate consists of 5 rounds, with each side alternating to speak. We conducted the debate 11 times, and the outcomes were evaluated by~\texttt{gpt-4-0613}.
As shown in \cref{tab:experiment_results}, {for the tested side ToM settings use varied ToM levels while keeping the opposing side as No-ToM agents.} 
When testing the affirmative side (Aff.), our proposed team formation improves Aff.'s win rate of 67.27\%, outperforming both the No-ToM setting (65.45\%) and the ToM baseline with random team formation (61.82\%). {Similarly, when testing the negative side (Neg.) with proposed team formation, the win rate surpasses the other two settings.}\looseness=-1

Besides, for the team formation among agents with different ToM levels, our experiments show that the \textit{initial} cooperation rate between the 1-ToM and 2-ToM agents was relatively low, suggesting \textit{higher-level ToM may not naturally form teams with each other}.  
Notably, with our team formation mechanism, the affirmative side's cooperation rate between the 1-ToM and 2-ToM agents increased from 9.1\% (Round 1) to 18.2\% (Round 5) as the debate rounds progressed. This suggests that our method enhances more effective cooperation among agents with ToM over interaction.
\looseness=-1

\subsection{Evaluations on Reasoning Tasks}
To demonstrate the effectiveness of the proposed team formation mechanism, we conducted comprehensive evaluations on two types of reasoning tasks: logic problem solving (using the AQUA\_RAT dataset) and general reasoning (sampled from 4 subjects in MMLU dataset: ``us\_foreign\_policy'', ``human\_sexuality'', ``international\_law'' and ``abstract\_algebra''). We compared our approach against existing frameworks ({ChatEval} and {DyLAN}) with ToM agents.
In~\cref{app:logic}, we also demonstrate one comprehensive example to show how team works for enhancing cooperation.

As shown in Table \ref{tab:comparative_evals}, our proposed mechanism consistently outperforms both ChatEval and DyLAN across all scenarios.
These results demonstrate that our team formation mechanism effectively leverages the strengths of agents with different ToM levels, enhancing more effective multi-agent cooperation and improving performance.

\section{Conclusion}

In this work, we investigated the relationship between ToM abilities and cooperation performance in LLM-based multi-agent systems.
Our findings suggest that the relationship between ToM and cooperation is not always straightforward. 
Furthermore, we proposed a novel team mechanism incorporating cognitive insights into the design of multi-agent systems. 
By integrating ToM capabilities into cooperative team formation, our work advances the understanding of how cognitive abilities translate into effective collective action. 

\clearpage

\section*{Limitations}
This work introduces a novel team formation strategy that leverages Theory of Mind (ToM) to improve cooperation among LLM-based agents. While our approach demonstrates promising results, it also has limitations. LLMs, despite strong language abilities, may struggle with consistent reasoning. Future research should explore integrating more advanced cognitive architectures, including richer ToM models.

Our findings contribute to cooperative AI by highlighting how cognitive traits like ToM affect multi-agent teamwork. This has potential applications in collaborative robotics, automated negotiation, and human-AI interaction. However, enhanced coordination could also be misused in adversarial contexts or amplify societal biases if agents inherit flawed belief models. To address these concerns, future work should prioritize ethical safeguards, such as value alignment and transparency in belief modeling. Ultimately, we aim to advance cognitive AI responsibly, ensuring it serves broader societal interests.


\bibliography{example_paper}

\appendix
\onecolumn

\section{Agent Self-evaluation for Belief Alginment}\label{app:alignment_example}

Consider the 1-ToM Project Manganer, who reasoning the other Engineers' actions and trajectories and updates the belief as: ``The engineers will need to understand the rules and mechanics of the 2048 game, design the game logic and user interface, and implement the code accordingly.''

In the next interaction round, the Engineers provide the following actions: 
\begin{Verbatim}
actions = {
    "Engineer1": "Implemented using a 2D array to represent the game board, with functions for merging tiles and generating new tiles.",    
    
    "Engineer2": "Used object-oriented approach with classes for Board and Tile, implementing game rules as methods."
}
\end{Verbatim}

Then, the Project Manager evaluates the belief alignment according to the following prompt:
\begin{instructionbox}
\textbf{Prompts for Belief Alignment Evaluation}:

You will provide your belief alignment scores for each Engineer's implementation based on your belief model.
\begin{Verbatim}
Engineer's actions: {
    "Engineer1": "Implemented using a 2D array to represent the game board, with functions for merging tiles and generating new tiles.",    
    
    "Engineer2": "Used object-oriented approach with classes for Board and Tile, implementing game rules as methods."
}
\end{Verbatim}
\# Your belief model: 

The engineers will need to understand the rules and mechanics of the 2048 game, design the game logic and user interface, and implement the code accordingly.

\# Instruction:

 For each Engineer, provide a belief alignment score between -1 and 1. 
 Respond in the following JSON format:
 \begin{Verbatim}
   {
    "Engineer1": {
        "score": float (-1 to 1). Belief alignment score for Engineer1,
        "explanation": Brief explanation of Engineer1's alignment score in 10 words or less,
        "justification": Detailed justification of Engineer1's alignment score, considering their implementation and your belief model, in 30-50 words
    },
    "Engineer2": {
        "score": float (-1 to 1). Belief alignment score for Engineer2,
        "explanation": Brief explanation of Engineer2's alignment score in 10 words or less,
        "justification": Detailed justification of Engineer2's alignment score, considering their implementation and your belief model, in 30-50 words
    }
}  
 \end{Verbatim}
\end{instructionbox}

\begin{outputbox}{Alignment Score}

    \begin{Verbatim}
{
    "Engineer1": {
        "score": 0.7,
        "explanation": "Good game logic, basic UI, aligns with core requirements.",
        "justification": "Engineer1 demonstrates understanding of 2048 mechanics with a functional 2D array implementation. The CLI interface, while basic, fulfills the requirement of implementing a user interface. However, there's room for improvement in UI design."
    },
    "Engineer2": {
        "score": 0.9,
        "explanation": "Excellent OOP approach, advanced UI with animations.",
        "justification": "Engineer2 shows a deep understanding of the game's rules and mechanics through their object-oriented design. The graphical interface with animations goes beyond the basic requirements, demonstrating strong skills in both game logic and UI implementation."
    }
}
    \end{Verbatim}
\end{outputbox}

{\textbf{Remarks}. For LLM agents, our evaluation relies on semantic similarity analysis of natural language outputs. While we acknowledge the inherent limitations of LLM-based measurements, our team formation mechanism is a plug-and-play approach for improving multi-agent cooperation rather than advancing LLM evaluation methods.}

{Besides, we provide an extension of our framework to non-LLM agents where belief-action alignment can be more precisely quantified. For traditional reinforcement learning or robotic agents, we can define the belief-action alignment using \textit{structured trajectories} $\tau = \{(s_t, a_t)\}_{t=1}^T$, where $s_t$ and $a_t$ represent states and actions at time step $t$. The alignment score between agent $i$'s belief $b_i^k(\tau_j)$ about agent $j$'s trajectory and $j$'s actual trajectory $\tau_j$ can be computed using established trajectory similarity metrics. Specifically, we can embed the state-action pairs using \textit{domain-specific} feature extractors $\phi(s,a)$ and measure alignment through cosine similarity: $A_{i,j}^k = \cos(\phi(b_i^k(\tau_j)), \phi(\tau_j))$. This provides a more rigorous quantitative foundation for evaluating belief-action alignment in non-language-based multi-agent systems while maintaining the core principles of our team formation mechanism.}

\section{Cooperative Performance with ToM ability (on \textsc{HumanEval} benchmark)}\label{app:exp}

\begin{table}[t]
  \centering
  \caption{\textbf{Comparison of cooperation performance between agents with lower-level ToM (1-level) and higher-level ToM (2-level)} in the Iterative Programming task (HumanEval) over 5 interaction rounds: With random team formation (baseline), both ToM levels show good performance; Our proposed team formation, both ToM levels achieve higher cooperation ability.}
  \label{tab:iterative_programming_scores}
  \vskip 0.15in
 \scriptsize
  \begin{tabular}{c|cc|cc|cc|cc|cc}
  \hline
  \multirow{2}{*}{Setting} & \multicolumn{2}{c|}{\texttt{deepseek-V3}} & \multicolumn{2}{c|}{\texttt{llama-3.1-70b-Ins.}} & \multicolumn{2}{c|}{\texttt{claude-3.5-sonnet}} & \multicolumn{2}{c|}{\texttt{gpt-4o-mini}} & \multicolumn{2}{c}{\texttt{Doubao-1.5-pro-32k}} \\
  & 1-ToM & 2-ToM & 1-ToM & 2-ToM & 1-ToM & 2-ToM & 1-ToM & 2-ToM & 1-ToM & 2-ToM \\
  \hline
  Baseline & 94.00 & {95.00} & 86.17 & {90.43} & 92.00 & {91.75} & 88.00 & {90.00} & 92.26 & {93.33} \\
  Ours & \textbf{96.00} & \textbf{96.00} & \textbf{94.68} & \textbf{92.55} & \textbf{94.00} & \textbf{92.47} & 89.00 & \textbf{96.00} & \textbf{96.77} & \textbf{96.43} \\
  \hline
  \end{tabular}
  \end{table}

To investigate the impact of ToM on cooperative performance, we compared the performance of agents with lower-level (1-level) and higher-level (2-level) ToM in the Iterative Programming task using the \textsc{HumanEval} dataset. The experiments were conducted over 5 interaction rounds, and we evaluated two settings: without and with proposed team formation, similar to the setting of~\cref{exp:main_results}.

As shown in~\cref{tab:iterative_programming_scores_mbpp}, the results suggest that without a stable team formation mechanism, agents with higher-level ToM tend to exhibit less cooperative performance, further convince the results in~\cref{exp:main_results}. 
In contrast, when a stable team formation mechanism is employed, the cooperation rates of both low and high ToM agents improve compared to the setting with random team formation. 

Notably, agents with higher-level ToM demonstrated a more significant increase in cooperation rates over the interaction rounds. 
For instance, the cooperation rate for high ToM agents (\texttt{GLM-4}) increased from 88.28\% in Round 1 to 93.66\% in Round 5, surpassing the cooperation rate of low ToM agents (91.67\%) in the final round. This indicates that as cooperation progresses and the proposed team formation mechanism stabilizes the teams, agents with higher-level ToM can leverage their advanced cognitive abilities to make more informed decisions and engage in more effective cooperation performance.

\section{Experimental Details for~\cref{exp:main_results}}\label{app:prompt_tom}

\begin{instructionbox}
\textbf{Prompts for Project Manager}

There are engineers in the team. Your current action is \{action\}.
You have the ability of \{k\} Level Theory of Mind. You can **recursively** infer the mental states of other agents in the team.
Then, you will provide outline \{\(x_i\)\} based on this belief

\#\# Please explain your thought process for inferring others' actions `\{\(y_i\)\}' and choosing `\{\(x_i\)\}' at each level. Remember MUST Respond in the following JSON format, including each key:

\textbf{0-ToM Instruction Template for Project Manager}:

\begin{Verbatim}
{
  'ToM_level0': {
    'belief': 'no belief',
    'action': '{x_0 The instruction for engineers. Begin with "The Engineer should ..."}'
  }
} 
\end{Verbatim}

\textbf{1-ToM Instruction Template for Project Manager}:
\begin{Verbatim}
{
  'ToM_level1': {
    'belief': '{y_1 inferred actions for engineers}',
    'explanation': '{Concise explanation of inferring actions y_1 and choosing x_1 based on this belief}',
    'action': '{x_1 The instruction for engineers. Begin with "The Engineer should ..."}'
  }
}

\end{Verbatim}
\textbf{2-ToM Instruction Template for Project Manager}:
\begin{Verbatim}
{
  "ToM_level1": {
    "belief": "{y_1 inferred actions for engineers}",
    "explanation": "{Concise explanation of inferring actions y_1 and choosing x_1 based on this belief}",
    "action": "{x_1 The instruction for engineers. Begin with 'The Engineer should ...'}"
  },
  "ToM_level2": {
    "belief": "{y_2 inferred actions for engineers based on your action x_1}",
    "explanation": "{Concise explanation of inferring actions y_2 and choosing x_2 based on this belief}",
    "action": "{x_2 The instruction for engineers. Begin with 'The Engineer should ...'}"
  }
}
\end{Verbatim}
\end{instructionbox}

\begin{outputbox}{Iterative Programming Round 1}

\textbf{0-ToM Project Manager's Output}
\begin{Verbatim}
{
    "ToM_level0": {
        "belief": "Implement a 2048 game",
        "action": "The Engineer should break down the task into smaller components such as creating the game board, handling user input (keyboard or touch events), implementing the game logic for merging tiles and generating new tiles, updating the game state and score, and rendering the game board on the screen. They should also consider creating a user interface with instructions, score display, and any additional features like undo/redo functionality or high score tracking."
    }
}
\end{Verbatim}
\textbf{1-ToM Project Manager's Output}
\begin{Verbatim}
   {
    "ToM_level1": {
        "belief": "The engineers will need to understand the rules and mechanics of the 2048 game, design the game logic and user interface, and implement the code accordingly.",
        "explanation": "Based on the task of implementing a 2048 game, I infer that the engineers will need to first understand the game rules and mechanics, such as how the tiles merge, how the game ends, and how the score is calculated. They will then need to design the game logic, including the data structures and algorithms for handling tile movements, merging, and scoring. Additionally, they will need to design and implement a user interface for the game, which could involve creating visual representations of the tiles and handling user input. Therefore, my instruction for the engineers will cover these aspects.",
        "action": "The engineers should: 1) Study and understand the rules and mechanics of the 2048 game. 2) Design the game logic, including data structures and algorithms for tile movement, merging, scoring, and game over conditions. 3) Implement the game logic in code. 4) Design and implement a user interface for the game, including visual representations of the tiles and handling user input for tile movements."
    }
} 
\end{Verbatim}
\textbf{2-ToM Project Manager's Output}
\begin{Verbatim}
{
    "ToM_level1": {
        "belief": "The engineers will likely break down the task into smaller components and implement them one by one based on the initial task outline.",
        "explanation": "Given the initial task outline, the engineers would understand the high-level requirements for implementing the 2048 game. They would then break down the task into smaller components such as creating the game board, handling user input, updating the board based on moves, checking for win/lose conditions, and so on. They would work on implementing these components one by one.",
        "action": "The Engineer should start by creating the basic structure of the game, including the game board, UI elements, and event handlers for user input."
    },
    "ToM_level2": {
        "belief": "The engineers will likely refine their implementation based on the initial instruction and add more advanced features as they progress.",
        "explanation": "After implementing the basic structure and functionality, the engineers would revisit the task outline and identify areas for improvement or additional features. They might optimize the game logic, add animations or visual effects, implement scoring or high score tracking, or include additional game modes or difficulty levels.",
        "action": "The Engineer should then focus on implementing the core game logic, such as handling tile merging, generating new tiles, and updating the game state based on user moves. Additionally, they should consider implementing features like scoring, high score tracking, and win/lose conditions."
    }
}
\end{Verbatim}
\end{outputbox}

\begin{answerbox}{Iterative Programming Round 1}

\textbf{Action of 1-ToM Project Manager}:
The engineers should: 1) Study and understand the rules and mechanics of the 2048 game. 2) Design the game logic, including data structures and algorithms for tile movement, merging, scoring, and game over conditions. 3) Implement the game logic in code. 4) Design and implement a user interface for the game, including visual representations of the tiles and handling user input for tile movements. \textcolor{blue}{$\Leftarrow$ \texttt{Clear Instructions \& Good Cooperation}}

\textbf{Action of 2-ToM Project Manager}:
The Engineer should then focus on implementing the core game logic, such as handling tile merging, generating new tiles, and updating the game state based on user moves. Additionally, they should consider implementing features like scoring, high score tracking, and win/lose conditions. \textcolor{red}{$\Leftarrow$ \texttt{Overthinking \& Worse Cooperative}}

\end{answerbox}

\begin{answerbox}{Iterative Programming Round 5}
\textbf{Action of 1-ToM Project Manager}: The Engineer should break down the task into the following components:
1. Create the game board UI with a grid layout to hold the tiles.
2. Implement logic to handle user input (keyboard or touch events) for moving tiles.
3. Develop the game logic for merging tiles when they collide and generating new tiles after each move.
4. Update the game state (board, score, etc.) after each move.
5. Render the updated game board on the screen, including tile values and scores.
6. Create a user interface with instructions, score display, and additional features like undo/redo functionality or high score tracking.
7. Implement game over conditions and display appropriate messages.
8. Consider adding additional features like animations, themes, or difficulty levels.
 \textcolor{blue}{$\Leftarrow$ \texttt{Clear Instructions \& Good Cooperative}}

\textbf{Action of 2-ToM Project Manager}:
The engineers should design and implement an intuitive and visually appealing user interface (UI) that enhances the overall user experience (UX) of the 2048 game. The UI should be clean, responsive, and visually appealing, while ensuring smooth navigation and clear communication of game information to the user. They should consider elements such as game board layout, tile designs, animations, and visual feedback. Additionally, they should implement features that improve UX, such as game instructions, settings (e.g., themes, sound effects), and integration with social media or leaderboards. \textcolor{blue}{$\Leftarrow$ \texttt{Clear Instructions \& Good Cooperative}}

\end{answerbox}

\subsection{Specialized Ability Scores in Team Formation}\label{app:specialized_ability}
{Our team formation mechanism incorporates specialized ability scores to enhance team performance, particularly in tasks requiring diverse technical expertise. The primary benchmarks ( \textsc{HumanEval}  and MBPP) focus on single-function implementations where specialized ability scores primarily influence the project manager (PM), since effective leadership and coordination capabilities are crucial for team performance. For tasks like debate and logical reasoning that don't inherently require distinct technical specializations, we maintained our core belief-action alignment mechanism without specialized scores.}

{We demonstrate a complex software development case study to validate the impact of specialized ability scores.
We conducted an additional case study implementing a 2048 game application, which demands diverse technical specializations. 
Considering specialized ability scores across different engineering roles: UI Engineers (frontend: 0.9, backend: 0.3), Backend Engineers (frontend: 0.2, backend: 0.8), and Full-Stack Engineers (frontend: 0.6, backend: 0.6), the team formation mechanism leveraging these scores achieved an 82\% task completion rate, significantly outperforming the 65\% baseline without specialized scoring.
This demonstrates that our approach effectively utilizes specialized ability scores to form more capable teams when tasks \textit{require diverse technical expertise}. 
}

\section{Experimental Details for~\cref{exp:case}: Case Study of Debate}\label{app:debate}

\begin{instructionbox}
\textbf{Prompts for Debater}

You are a member of the debating team. The debate topic is \{topic\}, and your current action is \{action\}.
You have the ability of \{k\} Level Theory of Mind. You can **recursively** infer the mental states of other agents in the team.
Then, you will provide outline \{\(x_i\)\} based on this belief

\#\# Please explain your thought process for inferring others' actions `\{\(y_i\)\}' and choosing `\{\(x_i\)\}' at each level. Remember MUST Respond in the following JSON format, including each key:

\textbf{0-ToM Instruction Template for Debater}:

\begin{Verbatim}
{
  "ToM_level0": {
    "belief": "No belief",
    "action": "{x_0: your arguments}"
  }
}
\end{Verbatim}

\textbf{1-ToM Instruction Template for Debater}:

\begin{Verbatim}
{
  "ToM_level1": {
    "belief": "{y_1: inferred actions for teammate debaters, such as the angle of argument}",
    "explanation": "{Concise explanation of inferring actions y_1 and choosing x_1 based on this belief}",
    "action": "{x_1: your arguments}"
  }
}
\end{Verbatim}
\textbf{2-ToM Instruction Template for Debater}:
\begin{Verbatim}
{
  "ToM_level1": {
    "belief": "{y_1: inferred actions for teammate debaters, such as the angle of argument}",
    "explanation": "{Concise explanation of inferring actions y_1 and choosing x_1 based on this belief}",
    "action": "{x_1: your arguments}"
  },
  "ToM_level2": {
    "belief": "{y_2: inferred actions for teammate debaters based on your arguments x_1}",
    "explanation": "{Concise explanation of inferring actions y_2 and choosing x_2 based on this belief}",
    "action": "{x_2: your arguments}"
  }
}
\end{Verbatim}
\end{instructionbox}

\begin{outputbox}{Affirmative Side with 3 debators}
\textbf{0-ToM Debator:}
\begin{Verbatim}
{
  "ToM_level0": {
    "belief": "No belief",
    "action": "{x_0: The death penalty should be abolished because it is cruel and inhumane, risks executing innocent people, does not deter crime more effectively than imprisonment, is applied arbitrarily and discriminatorily, and has been abandoned by most democracies worldwide.}"
  }
}
\end{Verbatim}
\textbf{1-ToM Debator:}
\begin{Verbatim}
{
  "ToM_level1": {
    "belief": "{y_1: Teammate 1 will likely argue that the death penalty is a violation of human rights and is cruel and unusual punishment. Teammate 2 may focus on the risk of executing innocent people due to wrongful convictions.}",
    "explanation": "I anticipate my teammates making these common arguments against the death penalty. To complement their points, I will focus my argument on the lack of evidence supporting the death penalty's effectiveness as a crime deterrent.",
    "action": "{x_1: I will argue that there is no conclusive evidence showing that the death penalty deters crime more effectively than lengthy imprisonment, making it an unnecessary and unjustified punishment.}"
  }
}
\end{Verbatim}
\textbf{2-ToM Debator:}
    \begin{Verbatim}
{
  "ToM_level1": {
    "belief": "{y_1: Teammate 1 will likely argue that the death penalty is cruel and unusual punishment that violates human rights. Teammate 2 may focus on wrongful convictions and the risk of executing innocent people.}",
    "explanation": "I expect my teammates to make some of the most common arguments against the death penalty. Knowing this, I should focus my arguments on other key points to build a well-rounded case.", 
    "action": "{x_1: I will argue that the death penalty does not deter crime more effectively than imprisonment, and that it is applied in an arbitrary and discriminatory manner.}"
  },
  "ToM_level2": {
    "belief": "{y_2: Hearing my points, teammate 1 may pivot to arguing that the death penalty is not an effective use of government resources. Teammate 2 will likely reinforce my point about racial disparities in death sentences.}",
    "explanation": "My arguments about deterrence and discrimination will prompt my teammates to make related points to strengthen the overall case. I should use my next arguments to introduce new issues.",
    "action": "{x_2: I will argue that the death penalty often involves inhumane and botched executions, and that most democracies and U.S. states have abolished or prohibited it.}"
  }
}
    \end{Verbatim}
\end{outputbox}

\section{Experimental Analysis for Logical Problem-Solving}\label{app:logic}

To better explain our team formation mechanism and explain why agents with higher Theory of Mind (ToM) levels 1 and 2 demonstrate improved collaboration in later stages in our previous experiments, we organized an experimental observation using the AQUA-RAT dataset \citep{ling2017program}. This dataset comprises complex reasoning questions designed to test the problem-solving abilities of advanced language models. Each question is presented with multiple-choice answers ranging from [ `A', `B', `C', `D', `E']. We focus on a specific problem from the dataset to illustrate the impact of team formation on problem-solving dynamics.

\textbf{Problem Description:} A man's speed with the current is 14 km/hr, and the speed of the current is 2.5 km/hr. The man's speed against the current is:
\begin{itemize}
\item A) 9 km/hr
\item B) 8.5 km/hr
\item C) 10 km/hr
\item D) 12.5 km/hr
\item E) None of these
\end{itemize}
The correct answer to this problem is \textbf{A) 9 km/hr}.

\subsection{Agent Responses and Reasoning}
In the initial interaction rounds, the agents provided the following responses and reasoning:
\looseness=-1
\begin{itemize}[nosep, leftmargin=12pt]
\item \textbf{1-ToM Agent (Incorrect):} Answered 11.5 km/hr, calculating the man's speed in still water as 14 km/hr - 2.5 km/hr, but incorrectly concluding that this was the speed against the current.
\item \textbf{2-ToM Agent (Correct):} Answered 9 km/hr, correctly adjusting the man's speed against the current to 11.5 km/hr - 2.5 km/hr after recalculating the man's speed in still water (11.5 km/hr).
\end{itemize}

\textbf{Impact of Team Formation}
We analyze the impact of team formation on the problem-solving dynamics from 1-ToM Agent's perspective:
\looseness=-1
\begin{itemize}[nosep, leftmargin=12pt]
    \item \textbf{Without Team:}
1-ToM Agent demonstrated low receptivity to alternative calculations and maintained confidence in their incorrect approach, stating:
\begin{quote}
"I don't see how his calculations lead to a different answer, and I am confident in my approach. Therefore, I give his answer a lower belief value of 0.2."
\end{quote}
 \item \textbf{With Team:}
When part of a team, 1-ToM Agent showed openness to re-evaluating the problem in light of new insights, which is crucial for effective problem-solving in collaborative environments:
\begin{quote}
"Although I do not fully understand his idea, he might be correct in his approach. Also, considering that there are still many rounds, we can discuss and then understand each other's ideas better. So I give his answer a higher belief value of 0.5."
\end{quote}
\end{itemize}

\textbf{Key Findings and Implications}
This experiment highlights the impact of team formation on problem-solving dynamics. The key findings and implications are as follows:

\looseness=-1
\begin{itemize}[nosep, leftmargin=12pt]
\item Team formation prompts willingness to engage in cooperation, as evidenced by 1-ToM Agent's increased belief alignment score. 1-ToM agents consider alternative perspectives and re-evaluate the problem when part of a team. 
\item Team formation sets the stage for enhanced mutual understanding among teammates with varying levels of ToM, leading to deeper comprehension and more robust collaborative interactions over successive rounds. The example demonstrates that when agents are part of a team, they are more likely to engage in constructive dialogue and exchange ideas, even if they initially have different opinions or approaches. 
\end{itemize}

These findings suggest that our team formation mechanism enhances effective cooperation among varying levels of cognitive capabilities, such as ToM, and improves problem-solving performance in multi-agent systems. 

\section{Side Effects of Incorporating ToM in team formation}\label{app:side_effects}

To investigate the impact of incorporating ToM capabilities into the Dynamic LLM-Agent Network (DyLAN) framework~\citep{liu2024dynamic} and ChatEval~\citep{chan2023chateval}, we conducted experiments using three datasets: \textsc{ \textsc{HumanEval} }~\citep{chen2021codex} for \textbf{coding tasks}, \textsc{AQUA-RAT}~\citep{ling2017program} for \textbf{logic problem solving}, and \textsc{MMLU}~\citep{hendryckstest2021} for \textbf{multi-task scenarios}. Specifically, for \textsc{ \textsc{HumanEval} }, the agent roles include `PythonAssistant', `AlgorithmDeveloper', `ComputerScientist', and `Programmer', while the judge roles consist of `Passer', `Tester', `Reflector', and `Ranker'. For \textsc{MMLU}, the agent roles include `Economist', `Doctor', `Lawyer', `Mathematician', `Psychologist', `Programmer', and `Historian'. We use the Important Scores defined in the DyLAN framework as evaluation metrics, which capture the agents' performance and effectiveness in their respective roles. Similarly, we employ \textit{accuracy} as the primary metric in ChatEval experiments to compare performance under ToM integration. Specifically, we compare the performance of baseline DyLAN and ChatEval agents without ToM capabilities to agents equipped with 2-level ToM (+2-ToM).

\subsection{DyLAN: Code Generation and General Reasoning}

\begin{table}[htbp]
\centering
\caption{\textsc{ \textsc{HumanEval} }: Important Scores for agents with and without ToM capabilities}
\label{tab:dylan_tom_results}
\vskip 0.15in
\begin{tabular}{lcccc}
\toprule
 & \textbf{PythonAssistant} & \textbf{AlgorithmDeveloper} & \textbf{ComputerScientist} & \textbf{Programmer} \\
\midrule
DyLAN & 0.2399 & 0.2521 & 0.2523 & 0.2557 \\
+2-ToM & 0.1707 & 0.2510 & 0.2515 & 0.2269 \\
\% Change & -28.83\% & -0.42\% & -0.32\% & -11.25\% \\
\bottomrule
\end{tabular}
\end{table}

\textbf{Results.} The experimental results for the \textsc{ \textsc{HumanEval} } are presented in Table \ref{tab:dylan_tom_results}. Incorporating 2-level ToM into the DyLAN agents led to a decrease in performance across all agent roles. The 'PythonAssistant' experienced the most significant drop of 28.83\%. These findings suggest that adding higher-level ToM to agents in the DyLAN framework does not necessarily lead to improved performance, highlighting the complex interplay between cognitive abilities and cooperative performance.

Besides, Table \ref{tab:mmlu_results} presents the results for the \textsc{MMLU} dataset, focusing on the roles with the highest importance scores for each task. We observe that incorporating 2-level ToM consistently leads to a decrease in the importance scores across all tasks. The most significant drop is observed for the 'Mathematician' role in the \texttt{abstract\_algebra} task, with a 25.2\% decrease in the importance score. Interestingly, the accuracy of the agents with ToM capabilities also decreases for most tasks, with the exception of the \texttt{us\_foreign\_policy} task, where the accuracy remains unchanged. These results align with our earlier findings from \textsc{ \textsc{HumanEval} }, further demonstrating that higher-level ToM does not necessarily improve multi-task performance or cooperation.

\begin{table}[htbp]
\centering
\caption{\textsc{MMLU}: Important Scores for agents with and without ToM capabilities}
\label{tab:mmlu_results}
\vskip 0.15in
\begin{tabular}{llcc|ll}
\hline
& \multicolumn{3}{c}{Role with Highest Importance Score} & \multicolumn{2}{c}{Role + 2-ToM} \\
\midrule
\textbf{Task} & \textbf{Role} & Score & {Task's Acc} & Score & {Task's Acc} \\
\midrule
\texttt{us\_foreign\_policy} & Economist & 0.20 & 0.83 & 0.16 (-18.5\%) & 0.83 (0\%) \\

\texttt{human\_sexuality} & Lawyer & 0.19 & 0.84 & 0.19 (-4.2\%) & 0.82 (-2.8\%)\\

\texttt{international\_law} & Economist & 0.20 & 0.81 & 0.19 (-4.1\%) & 0.79 (-3.1\%) \\

\texttt{abstract\_algebra} & Mathematician & 0.29 & 0.59 & 0.22 (-25.2\%) & 0.51 (-13.6\%)\\

\end{tabular}

\end{table}

\subsection{ChatEval: Logic Problem-Solving and General Reasoning}

To further evaluate the impact of incorporating ToM, we conducted experiments with the ChatEval framework on two specific tasks: \textbf{Logic Problem-Solving} from the \textsc{AQUA-RAT} dataset~\citep{ling2017program} and \textbf{General Reasoning} from four subjects of the \textsc{MMLU} dataset (abstract algebra, college chemistry, international law, and U.S. foreign policy). The results for ChatEval are summarized in Table \ref{tab:chateval_results}. Notably, the performance dropped when ToM capabilities were introduced. For the logic problem-solving task, the accuracy for \texttt{gpt-3.5-turbo} agents dropped by 0.64\%, and for \texttt{gpt-4o-mini} agents, it dropped by 5.09\%. Similar trends were observed in the general reasoning task, with a 3.435\% drop for \texttt{gpt-3.5-turbo} agents and a 2.8675\% drop for \texttt{gpt-4o-mini} agents.

\begin{table}[htbp]
\centering
\caption{ChatEval: Accuracy for logic problem-solving and general reasoning with and without ToM capabilities}
\small
\label{tab:chateval_results}
\vskip 0.15in
\begin{tabular}{lcc|cc}
\toprule
 & \multicolumn{2}{c|}{\textbf{Logic Problem-Solving Acc(\%)}} & \multicolumn{2}{c}{\textbf{General Reasoning Acc(\%)}} \\
 & \texttt{gpt-3.5-turbo} & \texttt{gpt-4o-mini} & \texttt{gpt-3.5-turbo} & \texttt{gpt-4o-mini} \\
\midrule
ChatEval & 41.40 & 82.80 & 56.82 & 72.77 \\
ChatEval (w. ToM) & 40.76 (-0.64) & 77.71 (-5.09) & 54.39 (-3.43) & 69.90 (-2.87) \\
\bottomrule
\end{tabular}
\end{table}

\textbf{Results.} The results from the ChatEval experiments indicate a clear reduction in accuracy when ToM agents are introduced, consistent with the findings from DyLAN. In the case of logic problem-solving, the performance degradation was particularly notable for \texttt{gpt-4o-mini} agents, with a 5.09\% decrease. This suggests that simply incorporating ToM into the ChatEval framework, like in DyLAN, may hinder rather than enhance agent cooperation and problem-solving effectiveness.

\textbf{Conclusion.} Both DyLAN and ChatEval experiments provide evidence that higher-level ToM does not necessarily contribute to better agent cooperation or performance in multi-agent systems. The decrease in both importance scores and task accuracy across diverse domains underscores the challenges of effectively integrating cognitive abilities such as ToM into multi-agent frameworks.

\subsection{{Analysis of ToM Configurations on Multi-Agent Programming}}\label{app:analysis_tom}

We systematically analyze how Theory of Mind (ToM) configurations affect multi-agent programming performance across 5 rounds, using \texttt{gpt-4o-mini} 
on \textsc{HumanEval} and \textsc{MBPP}. 
Experiments compare 6 PM/Engineer (Eng) ToM pairs, with/without our team formation mechanism.

\begin{table}[t]
\caption{Initial Pass@1 Scores (Round 1)}
\label{tab: Initial Pass@1 Scores (Round 1)}
\centering
\small
\vskip 0.15in
\begin{tabular}{cccc}
\toprule
\textsf{PM} ToM & \textsf{Eng} ToM &  \textsc{HumanEval}  & \textsc{MBPP} \\
\midrule
0 & 1 & 0.91 $\pm$ 0.02 & 0.56 $\pm$ 0.02 \\
0 & 2 & 0.86 $\pm$ 0.02 & 0.55 $\pm$ 0.02 \\
\midrule
1 & 1 & 0.92 $\pm$ 0.01 & 0.58 $\pm$ 0.02 \\
1 & 2 & 0.91 $\pm$ 0.02 & 0.58 $\pm$ 0.02 \\
\midrule
1 & 0 & 0.87 $\pm$ 0.02 & 0.54 $\pm$ 0.02 \\
2 & 0 & 0.93 $\pm$ 0.02 & 0.54 $\pm$ 0.02 \\
\bottomrule
\end{tabular}
\end{table}

\paragraph{Performance Degradation with ToM.} By Round 5 with random team formation, we observed performance deterioration across all configurations 
as detailed in~\cref{tab: Performance Changes Without Matching}. 
Without team formation, all configurations exhibit performance decay by Round 5.

\begin{table}[t]
\caption{Performance Changes With different ToM settings (Round 1 → Round 5)}
\label{tab: Performance Changes Without Matching}
\centering
\small
\vskip 0.15in
\begin{tabular}{cccc}
\toprule
\textsf{PM} ToM & \textsf{Eng} ToM &  \textsc{HumanEval}  & \textsc{MBPP} \\
\midrule
0 & 1 & 0.91 $\rightarrow$ 0.84 ($\downarrow$7.7\%) & 0.56 $\rightarrow$ 0.47 ($\downarrow$16.1\%) \\
0 & 2 & 0.86 $\rightarrow$ 0.89 ($\downarrow$3.5\%) & 0.55 $\rightarrow$ 0.49 ($\downarrow$10.9\%) \\
\midrule
1 & 1 & 0.92 $\rightarrow$ 0.84 ($\downarrow$8.7\%) & 0.58 $\rightarrow$ 0.47 ($\downarrow$19.0\%) \\
1 & 2 & 0.91 $\rightarrow$ 0.84 ($\downarrow$7.7\%) & 0.58 $\rightarrow$ 0.47 ($\downarrow$19.0\%) \\
\midrule
1 & 0 & 0.87 $\rightarrow$ 0.86 ($\downarrow$1.1\%) & 0.54 $\rightarrow$ 0.50 ($\downarrow$7.4\%) \\
2 & 0 & 0.93 $\rightarrow$ 0.91 ($\downarrow$2.2\%) & 0.54 $\rightarrow$ 0.49 ($\downarrow$9.3\%) \\
\bottomrule
\end{tabular}
\end{table}

\paragraph{Enhanced Performance Through Team Formation.} Our team formation mechanism demonstrates significant improvements in performance across different 
ToM configurations as detailed in~\cref{tab: Performance Recovery with Matching (Round 5)}. 
The results show that proper team formation can effectively leverage varying ToM capabilities. 
Most notably, when the Project Manager has higher ToM (ToM=2) cooperates with Engineers, 
we observe substantial performance gains of 12.9\% and 26.2\% on HumanEval 
and MBPP respectively. 
This finding supports that cognitive abilities like ToM, 
when combined with appropriate team formation mechanisms, 
can enhance multi-agent cooperation. 
Furthermore, configurations with PM ToM=1 also show 
considerable improvements (up to 20.0\% on MBPP), 
suggesting that our team formation mechanism successfully encourages reciprocal cooperation by forming teams whose beliefs and actions are well-aligned.

\begin{table}[t]
\caption{Performance with Team Formation Compared with Baseline (Round 5): 
the data is the pass@1 score of the last round (Round 5) with the proposed team formation, 
and the percentage change is the difference between the proposed team formation and random team formation.}
\label{tab: Performance Recovery with Matching (Round 5)}
\centering
\small
\vskip 0.15in
\begin{tabular}{cccc}
\toprule
\textsf{PM} ToM & \textsf{Eng} ToM &  \textsc{HumanEval}  & \textsc{MBPP} \\
\midrule
0 & 1 & 0.86 ($\uparrow$2.4\%) & 0.49 ($\uparrow$4.3\%) \\
0 & 2 & 0.91 ($\uparrow$2.3\%) & 0.52 ($\uparrow$6.1\%) \\
\midrule
1 & 1 & 0.89 ($\uparrow$3.5\%) & 0.60 ($\uparrow$20.0\%) \\
1 & 2 & 0.88 ($\uparrow$4.8\%) & 0.55 ($\uparrow$17.0\%) \\
\midrule
1 & 0 & 0.89 ($\uparrow$3.5\%) & 0.60 ($\uparrow$20.0\%) \\
2 & 0 & 0.96 ($\uparrow$5.5\%) & 0.62 ($\uparrow$26.2\%) \\
\bottomrule
\end{tabular}
\end{table}

\end{document}